\documentclass[prl,twocolumn]{revtex4}
\usepackage{graphicx}
\usepackage{bm}
\usepackage{subfigure}
\usepackage{amsmath}

\newcommand{\epl}{Europhys. Lett.~}

\newcommand{\pr}{Phys. Rev.~}
\newcommand{\pla}{Phys. Lett. A~}
\newcommand{\jpa}{J. Phys. A~}

\newcommand{\njp}{New J. Phys.~}
\newcommand{\nature}{Nature.~}

\newcommand{\e}{\mbox{e}}

\newcommand{\etals}{{\em et al.}}

\begin{document}

\title{Mesoscopic dynamical differences from quantum state preparation in a Bose-Hubbard trimer}

\author{M.~K. Olsen$^{1,2}$, T.W. Neely$^{1}$,  and A.S. Bradley$^{2}$}
\affiliation{$^{1}$School of Mathematics and Physics, University of Queensland, Brisbane, 
Queensland 4072, Australia.\\
$^{2}$Quantum Science Otago and Dodd-Walls Centre for Photonic and Quantum Technologies, Department of Physics, University of Otago, Dunedin 9016, New Zealand.\\
$^{3}$ARC Centre of Excellence for Engineered Quantum Systems, University of Queensland, Brisbane Queensland 4072, Australia.}

\date{\today}

\begin{abstract}

Conventional wisdom is that quantum effects will tend to disappear as the number of quanta in a system increases, and the evolution of a system will become closer to that described by mean field classical equations. In this letter we combine newly developed experimental techniques to propose and analyse an experiment using a Bose-Hubbard trimer where the opposite is the case. We find that differences in the preparation of a centrally evacuated trimer can lead to readily observable differences in the subsequent dynamics which increase with system size. Importantly, these differences can be detected by the simple measurements of atomic number.

\end{abstract}

\maketitle


Experimental techniques for the manipulation of ultracold bosons have recently approached the sophisticated level of control that have been available for many years with quantum optics. In particular, the high level of control of optical potentials~\cite{painting,BellPaint,DalibardDMD,Tylerpaint,Weitenberg} and the ability to address individual wells of a lattice system with an electron beam~\cite{NDC,Labouvie} have opened new vistas for cold atom experimentation. On the other hand, the exact quantum state of a trapped Bose-Einstein condensate (BEC) aroused early theoretical interest~\cite{Lewenstein,Dunningham}. Lewenstein and You~\cite{Lewenstein} used the Bogoliubov formalism to arrive at an amplitude squeezed state and an analytic expression for the phase diffusion of a trapped BEC.  Dunningham~\cite{Dunningham} went further and found that if Gaussian statistics were not assumed, the statistics became closer to those of a number squeezed state. They proposed that these states could be differentiated via measurements of phase diffusion.

In this letter we analyse a system where the difference between initial quantum states can be detected in the mean-field evolution, with no need for phase-sensitive measurements. Although such mean-field effects have been predicted in the photoassociation of atoms into 
molecules~\cite{BECstate1,BECstate2,BECFock}, this remains a difficult experiment to perform. The simpler mechanism we propose consists of an appropriately prepared three well Bose-Hubbard chain~\cite{BHmodel,Jaksch,BHJoel,Nemoto} with no initial occupation of the central well~\cite{Penna} and is similar to that previously investigated in terms of negative differential conductivity~\cite{NDCJoel}. We will show that the short time dynamics of the mean-fields are heavily affected by the choice of initial quantum state, and discuss how these initial states may be prepared experimentally.  


Our physical model consists of three inline potential wells in the tunnelling regime of the Bose-Hubbard model. This configuration has been analysed in terms of entanglement with the middle well initially 
occupied~\cite{BECsplit,splitOC}, for which the mean-field Fock and coherent state dynamics were indistinguishable. The effect of the quantum states was seen only in the quantum correlations used to measure entanglement. In this work, we consider that the two end wells are initially occupied with almost equal numbers of atoms. The quantum states of these occupied atomic modes will depend on how the system is prepared. If, for example, three full wells were prepared and the electron beam method of Labouvie \etals~\cite{NDC} were used to evacuate the central site, we may assume that the two occupied wells would be inhabited by something close to coherent or amplitude squeezed states. If, on the other hand, digital micromirror-based potentials~\cite{DalibardDMD,Tylerpaint,Weitenberg} were used to prepare two separate and isolated wells containing independently condensed atoms with negligible atom losses, and these were then translated together about a third well to begin the tunnelling, the initial states would be closer to Fock states~\cite{ZhangECS}. Using phase space methods for different quantum states~\cite{states}, we will investigate these three options.

Our description begins with the three-well inline Bose-Hubbard unitary interaction Hamiltonian,
\begin{equation}
{\cal H} = \hbar\chi\sum_{i=1}^{3}\hat{a}_{i}^{\dag\,2}\hat{a}_{i}^{2}-\hbar J \left(\hat{a}_{1}^{\dag}\hat{a}_{2}+\hat{a}_{2}^{\dag}\hat{a}_{1} +\hat{a}_{2}^{\dag}\hat{a}_{3}+\hat{a}_{3}^{\dag}\hat{a}_{2} \right),
\label{eq:genHam3line}
\end{equation}
where $\hat{a}_{i}$ is the bosonic annihilation operator for the atomic mode in the $i$th well, $\chi$ represents the collisional nonlinearity due to atomic s-wave scattering, and $J$ is the tunneling strength. To analyse the dynamics of a system described by this Hamiltonian, we prefer to use stochastic integration in the phase space representations. Our first preference is the positive-P representation~\cite{Pplus} which is exact where integration of the stochastic equations converges. 
The second is the truncated Wigner representation~\cite{Graham,Steel} which is approximate, but always converges for these type of systems. It also has the advantage that the probability distributions for the occupation of each well can be accurately calculated in the Bose-Hubbard model~\cite{BobWig}. In this work we will use the positive-P results for means and variances as an accuracy check of the truncated Wigner results.

We obtain the equations of motion by following the usual procedures~\cite{QNoise,DFW}, mapping the von Neumann equation for the density operator onto Fokker-Planck equations for the pseudoprobability functions. The positive-P representation gives a genuine Fokker-Planck equation while the Wigner representation results in a generalised Fokker-Planck equation with third order derivatives. Although this can be mapped onto stochastic difference equations, these are highly unstable~\cite{WigEPL}, so we take the usual route and truncate at second order, which is often justified as cutting the highest order in a $1/N$ expansion. 

\begin{figure}[tbhp]
\includegraphics[width=0.75\columnwidth]{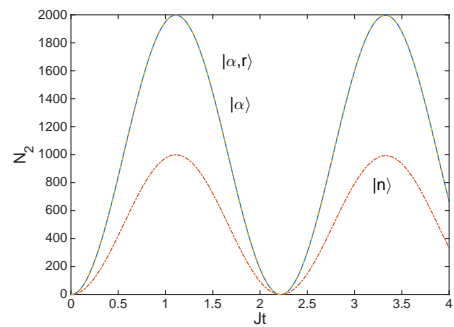}
\caption{(colour online) The number of atoms in the middle well as a function of scaled time $Jt$, for $N_{1}(0)=N_{3}(0)=1000$ and $\chi=10^{-4}$. The dash-dotted line is for initial number states and the upper, indistinguishable lines are for initial coherent and squeezed $(r=0.5)$ states. These results are the averaged solutions for $10^{6}$ trajectories of the positive-P equations and have converged to less than the plotted linewidths. All quantities in this and subsequent plots are dimensionless.}
\label{fig:N2ki4}
\end{figure}

For the positive-P representation, this process results in the It\^o calculus~\cite{SMCrispin} stochastic differential equations,
\begin{eqnarray}
\frac{d\alpha_{1}}{dt} &=& -2i\chi\alpha_{1}^{+}\alpha_{1}^{2}+iJ\alpha_{2} + \sqrt{-2i\chi\alpha_{1}^{2}}\;\eta_{1}, \nonumber \\
\frac{d\alpha_{1}^{+}}{dt} &=& 2i\chi\alpha_{1}^{+\,2}\alpha_{1}-iJ\alpha_{2}^{+} + \sqrt{2i\chi\alpha_{1}^{+\,2}}\;\eta_{2}, \nonumber \\
\frac{d\alpha_{2}}{dt} &=& -2i\chi\alpha_{2}^{+}\alpha_{2}^{2} + iJ\left(\alpha_{1}+\alpha_{3} \right) + \sqrt{-2i\chi\alpha_{2}^{2}}\;\eta_{3}, \nonumber \\
\frac{d\alpha_{2}^{+}}{dt} &=& 2i\chi\alpha_{2}^{+\,2}\alpha_{2} - iJ\left(\alpha_{1}^{+}+\alpha_{3}^{+} \right) + \sqrt{2i\chi\alpha_{2}^{+\,2}}\;\eta_{4}, \nonumber \\
\frac{d\alpha_{3}}{dt} &=& -2i\chi\alpha_{3}^{+}\alpha_{3}^{2} + iJ\alpha_{2} + \sqrt{-2i\chi\alpha_{3}^{2}}\;\eta_{5}, \nonumber \\
\frac{d\alpha_{3}^{+}}{dt} &=&2i\chi\alpha_{3}^{+\,2}\alpha_{3} - iJ\alpha_{2}^{+} + \sqrt{2i\chi\alpha_{3}^{+\,2}}\;\eta_{6},
\label{eq:PPequations}
\end{eqnarray}
where the $(\alpha_{j},\alpha_{j}^{+})$ are the c-number variables corresponding to the operators $(\hat{a}_{j},\hat{a}_{j}^{\dag})$ in the sense that the averages $\overline{\alpha_{j}^{m}\alpha_{k}^{+\,n}}$ converge in the limit over a large number of stochastic trajectories to the expectation values of normally-ordered operator products, $\langle\hat{a}_{k}^{\dag\,n}\hat{a}_{j}^{m}\rangle$. In general, $\alpha_{i}$ and $\alpha_{i}^{+}$ are not complex conjugates, with this freedom allowing us to reproduce quantum evolution. The $\eta_{j}$ are Gaussian random variables with the correlations $\overline{\eta_{j}(t)}=0$ and $\overline{\eta_{j}(t)\eta_{k}(t')}=\delta_{jk}\delta(t-t')$. These equations are solved numerically, taking averages over a large number of stochastic trajectories, with $10^{6}$ being averaged over for the results of Fig.~\ref{fig:N2ki4}. 

The use of the truncated Wigner representation results in the classical looking equations of motion,
\begin{eqnarray}
\frac{d\alpha_{1}}{dt} &=& -2i\chi |\alpha_{1}|^{2}\alpha_{1}+iJ\alpha_{2}, \nonumber \\
\frac{d\alpha_{2}}{dt} &=& -2i\chi |\alpha_{2}|^{2}\alpha_{2}+iJ\left(\alpha_{2}+\alpha_{3}\right), \nonumber \\
\frac{d\alpha_{3}}{dt} &=& -2i\chi |\alpha_{3}|^{2}\alpha_{3}+iJ\alpha_{2},
\label{eq:Wequations}
\end{eqnarray}
with quantum evolution resulting from different initial conditions sampled from distributions for the appropriate quantum 
states~\cite{states}. Averages of the Wigner variables represent symmetrically ordered operator products such that, for example $\overline{\alpha_{j}\alpha_{j}^{\ast}}$ converges to $\frac{1}{2}\langle \hat{a}_{i}^{\dag}\hat{a}_{i}+\hat{a}_{i}\hat{a}_{i}^{\dag} \rangle = N_{i}+\frac{1}{2}$. To obtain the binned probability distributions used below, we averaged over $10^{7}$ stochastic trajectories.


We begin our simulations with the outer two wells similarly populated with the same atomic quantum states, and investigate the effects of varying the collisional nonlinearity. 
The three states used, Fock, coherent, and amplitude squeezed, do not need equal numbers. We investigated differences of up to $10\%$ on each side, and found that this made little quantitative difference to the results. The results for coherent and amplitude squeezed states are first shown with the same mean numbers and phase, since these would most likely be prepared by evacuating the central well of a three-well system in equilibrium. The last result shown will demonstrate that having the same phase is not necessary to see the effects of interest. 

It is important that theory should be relevant to experiment and the parameters used here are consistent with accessible experimental values. Fixing the tunnelling rate at $J=1$ sets the scale for all the other parameters. $J$ itself can be altered by changes in the well depths and separation. The most difficult parameter to change experimentally would be $\chi$, which is possible using Feshbach resonance techniques~\cite{Feshbach}. Using the published results of Albiez \etals~\cite{Albiez} and setting their tunnelling rate equal to one, we find that their $\chi\approx 10^{-4}$ in our units. While this is the smallest value that we have used, deeper wells would lower $J$ and give a ratio $\chi/J$ consistent with our higher values, or $\chi$ could be changed using Feshbach techniques. By reference to the same article, we can also say that our system is in the regime where the three-mode approximation is valid.

 Fig.~\ref{fig:N2ki4} shows the mean field solutions for the middle well, for initial conditions of Fock, coherent, and squeezed states in the two outer wells with the middle well initially unpopulated. In all our calculations we set $J=1$, which sets the time scale. For the figure we have used the lower of three collisional nonlinearities considered, with $\chi=10^{-4}$. The coherent states $|\alpha\rangle$ are chosen with $\alpha_{1}=\alpha_{3}=\sqrt{10^{3}}$, the Fock states $|n\rangle$ are chosen with $n=1000$, and the amplitude squeezed states $|\alpha,r\rangle$ are chosen with $\alpha$ as in the coherent states and the squeezing parameter $r=0.5$. This means that the $\hat{X}(=\hat{a}+\hat{a}^{\dag})$ quadratures will have variances of $\e^{-r}\approx 0.6$ and the $\hat{Y}(=-i(\hat{a}-\hat{a}^{\dag}))$ quadratures will have $\e^{r}\approx 1.6$ at the beginning of the time evolution. The dynamics for the two other parameter regimes we have investigated, with $\chi=10^{-3}$ and $N_{1}(0)=N_{3}(0)=100$, as well as $\chi=10^{-2}$ with $N_{1}(0)=N_{3}(0)=20$, follow the same pattern. 
 
A similar difference was seen in a previous work investigating a four well phase sensitive gate for atoms~\cite{phasegate}, with the dynamics being markedly different for different initial quantum states. In that paper the difference in the dynamics was not quantified apart from showing the overlap or lack thereof of the mean field solutions plus and minus one standard deviation for the different initial conditions. Since the publication of that article, Lewis-Swan \etals~\cite{BobWig} have developed a method for binning the results of truncated Wigner trajectories and shown that, for broad enough number distributions, this can reproduce the actual number probability distribution, $P(N)$, to a high degree of accuracy. In particular, it was shown that the method worked well for the Bose-Hubbard dimer. 

\begin{figure}[tbhp]
\includegraphics[width=0.75\columnwidth]{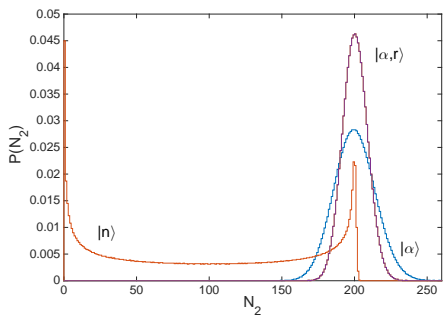}
\caption{(colour online) The number distributions from binning of truncated Wigner trajectories at the first time of maximum transfer, $Jt=1.11$, for $N_{1}(0)=N_{3}(0)=100$ and $\chi=10^{-3}$. The squeezed state has $r=0.5$.}
\label{fig:PNki3}
\end{figure}

\begin{figure}[tbhp]
\includegraphics[width=0.75\columnwidth]{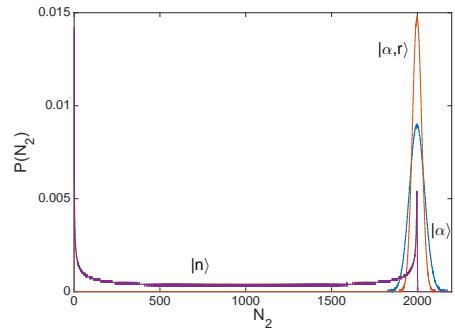}
\caption{(colour online) The number distributions from binning of truncated Wigner trajectories at the first time of maximum transfer, $Jt=1.11$,, for $N_{1}(0)=N_{3}(0)=1000$ and $\chi=10^{-4}$. The squeezed state has $r=0.5$.}
\label{fig:PNki4}
\end{figure}

We have used this method of binning of trajectories to calculate $P(N_{2})$ for the different initial conditions chosen. In Fig.~\ref{fig:PNki3} we show the result for $\chi=10^{-3}$. This obviously gives far more information about possible experimental outcomes than is available from the means and variances. We see that the distributions for initial coherent and squeezed states are similar, with that for the squeezed states being a little narrower. It would be difficult to tell these apart experimentally. The distribution for initial Fock states, however, is remarkably different, in both the mean and the distribution. Although the mean shows that the middle well will on average have $100$ atoms at this time of maximum transfer, the largest probability is for no atoms to be transferred at all. Many of the results of individual experimental runs would find a number that would be practically impossible when beginning with coherent states. Fig.~\ref{fig:PNki4} shows similar behaviour for $\chi=10^{-4}$ and $N_{1}(0)=N_{3}(0)=10^{3}$. In this case the most probable central well occupation for initial Fock states is again zero, and most of the possible numbers are outside the distributions for the coherent and squeezed states.

Although the pictorial results give a good indication of the difference between ensembles of outcomes, we feel that physics should be as quantitative as possible. 
To that end, we now move to a quantitative measure of the differences between predicted experimental outcomes, using a statistical method developed for the comparison of probability distributions.  
A measure of the difference between the ensembles of results can be given by the Bhattacharyya coefficient and the Bhattacharya statistical distance~\cite{Bhattacharyya}, which are statistical measures used to compare two distributions.  
The Bhattacharyya coefficient is defined for two discrete samples as
\begin{equation}
{\cal B}(P_{1},P_{2}) = \sum_{n=0}^{\infty}\sqrt{P_{1}P_{2}},
\label{eq:coefficient}
\end{equation}
and will be equal to one for identical distributions and zero when there is absolutely no overlap. The Bhattacharrya coefficents for the calculated distributions are given in the table below,

\begin{center}
 \begin{tabular}{|| c || c  || c   || c   ||} 
 \hline
   & $\chi =10^{-2}$ & $\chi=10^{-3}$ & $\chi=10^{-4}$  \\ [0.5ex] 
 \hline\hline 
 ${\cal B}(P_{F},P_{C})$ & $ 0.531 $ & $ 0.403 $ & $ 0.287 $ \\ 
 \hline
 ${\cal B}(P_{F},P_{S})$ & $ 0.484  $ & $ 0.364 $ & $ 0.259  $ \\ 
 \hline
 ${\cal B}(P_{C},P_{S})$ & $ 0.947 $ & $ 0.942 $ & $ 0.939 $ \\ 
 \hline
\end{tabular},
\end{center} 
\label{tab:Bcoeff}
where F=Fock, C=coherent, and S=squeezed with $r=0.5$.
From these we can calculate the Bhattacharyya statistical distance,
\begin{equation}
D_{AB} = -\ln\left[ {\cal B}(P_{A},P_{B})\right],
\label{eq:Bdiff}
\end{equation}
which has a value of zero for completely overlapping and hence indistinguishable distributions and becomes infinite for distributions with no overlap whatsoever. These are given in the following table,

\begin{center}
 \begin{tabular}{|| c || c  || c   || c   ||} 
 \hline
   & $\chi =10^{-2}$ & $\chi=10^{-3}$ & $\chi=10^{-4}$  \\ [0.5ex] 
 \hline\hline 
 $ D_{FC} $ & $ 0.633 $ & $ 0.909 $ & $ 1.25 $ \\ 
 \hline
 $ D_{FS} $ & $ 0.726  $ & $ 1.01 $ & $ 1.35  $ \\ 
 \hline
 $ D_{CS} $ & $ 0.055 $ & $ 0.060 $ & $ 0.063 $ \\ 
 \hline
\end{tabular}.
\end{center} 
\label{tab:Bdiff}
We immediately notice that the Bhattacharyya distance grows for each comparison as $\chi$ decreases. Given that a larger number of atoms can occupy each well, while staying within the confines of the Bose-Hubbard model, as long as $\chi$ decreases, this means that the difference between these distributions is a quantum statistical effect that becomes more observable as the number of atoms is increased. This example contradicts the conventional wisdom, where quantum effects are thought to disappear as the number of quanta increase, and a classical mean-field analysis is expected to become more accurate. The increase in the Bhattacharyya distance is directly due to the increased number of atoms and the quantum nature of the atomic field. Such an effect would be very difficult to observe with photons due to the difficulty of preparing large $N$ Fock states.
 
\begin{figure}[tbhp]
\includegraphics[width=0.75\columnwidth]{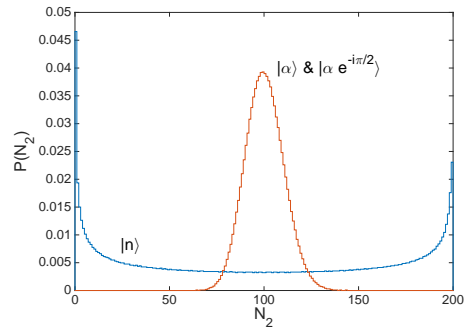}
\caption{(colour online) The number distributions from binning of truncated Wigner trajectories at the first time of maximum transfer, $Jt=1.11$,, for $N_{1}(0)=N_{3}(0)=100$ and $\chi=10^{-3}$. The wide distribution comes from initial Fock states, while the narrow distribution comes from coherent states with a $\pi/2$ initial phase difference.}
\label{fig:alphaphi}
\end{figure}

In previous work on a phase sensitive atomic gate constructed from a Bose-Hubbard model~\cite{phasegate}, we found that the differences in transferred populations due to initial states could be mimicked via phase differences between atomic modes. For example, two modes in coherent states with a $\pi$ phase difference resulted in an almost total suppression of tunnelling, while a  $\pi/2$ phase difference was found to give similar mean-field predictions to two Fock states. In the latter case, although the number variances were higher for two initial Fock states, this was not sufficient to conclusively differentiate the two different initial states. For the system we consider here, the same is true and either two Fock states or two coherent states with a $\pi/2$ phase difference evolve to give the same average number of atoms transferred. However, as can be seen in Fig.~\ref{fig:alphaphi}, the distributions leading to these averages are markedly different. This difference can again be quantified with ${\cal B}(P_{F},P_{\phi})=0.407$ and $D_{F\phi}=0.899$, where $P_{\phi}$ is now the distribution for the coherent states with $\pi/2$ phase difference.

In conclusion we have shown that newly developed techniques for the manipulation of trapped ultracold atoms allow for the observation of mesoscopic quantum effects using only number measurements. Previously the differences in behaviour of different quantum states have needed phase sensitive techniques such as homodyne measurement or coherence measurements in order to be detetcted. Here, a system as simple as a Bose Hubbard trimer can result in quantum dynamical effects which actually become easier to detect as the number of atoms is increased. This is the opposite of conventional wisdom and the realisation that these effects are not only possible but also readily observable has the potential to open new areas of investigation in quantum atom optics.

Experimentally, challenges remain in implementing the described Bose-Hubbard trimer system. However, dynamical optical control of sites in 1D lattices allows preparation of the separate wells with tunnelling rates similar to~\cite{Albiez}, with either electron beam emptying of lattice sites~\cite{NDC, Labouvie}, or with digital micromirror based methods to quickly empty the middle well~\cite{Weitenberg}. Deterministic preparation of Fock states with exact N is challenging, typically requiring a local number-sensitive energy shift and has been most successfully implemented in 3D optical lattice experiments~\cite{Greiner}. We note however that the large Bhattacharyya distances indicate results will be robust to shot-to-shot number fluctuations. Since BECs created in isolation will approximate Fock states in the limit of low atom losses~\cite{ZhangECS}, we expect the preparation step may be accomplished through deep evaporative cooling in potentials based on the aforementioned technologies.


\begin{thebibliography}{99}

\bibitem{painting}{K. Henderson, C. Ryu, C. MacCormick, and M.G. Boshier, \njp {\bf 11}, 043030 (2009).}
%
\bibitem{BellPaint}{T. A. Bell, J. A. P. Glidden, L. Humbert, M. W. J. Bromley, S. A. Haine, M. J. Davis, T. W. Neely, M. A. Baker and H. Rubinsztein-Dunlop, \njp {\bf18}, 089501 (2017).}
%
\bibitem{DalibardDMD}{J. L. Ville, T. Bienaim\'e, R. Saint-Jalm, L. Corman, M. Aidelsburger, L. Chomaz, K. Kleinlein, D. Perconte, S. Nascimb\'ene, J. Dalibard, and J. Beugnon, \pra {\bf 95}, 013632 (2017).}
%
\bibitem{Tylerpaint}{G. Gauthier, I. Lenton, N. McKay Parry, M. Baker, M. J. Davis, H. Rubinsztein-Dunlop, and T. W. Neely, Optica {\bf 3}, 1136 (2016).}
%
\bibitem{Weitenberg}{C. Weitenberg, M. Endres, J. F. Sherson, M. Cheneau, P. Schaub, T. Fukuhara, I. Bloch and S. Kuhr, \nature {\bf 471}, 319-324 (2011).}
%
\bibitem{NDC}{R. Labouvie, B. Santra, S. Heun, S. Wimberger, and H. Ott, \prl {\bf 115}, 050601 (2015).}
%
\bibitem{Labouvie}{R. Labouvie, B. Santra, S. Heun, and H. Ott, \prl {\bf 116}, 235302 (2016).}
%
\bibitem{Lewenstein}{M. Lewenstein and L. You, \prl {\bf 77}, 3489 (1996).}
%
\bibitem{Dunningham}{J.A. Dunningham, M.J. Collett, and D.F Walls, \pla {\bf 245}, 49 (1998).}
%
\bibitem{BECstate1}{M.K. Olsen and L.I. Plimak, \pra{\bf 68}, 031603 (2003).}
%
\bibitem{BECstate2}{M.K. Olsen, \pra {\bf 69}, 013601 (2004).}
%
\bibitem{BECFock}{M.K. Olsen, A.S. Bradley, and S.B. Cavalcanti, \pra {\bf 70}, 033611 (2004).}
%
\bibitem{BHmodel}{H. Gersch and G. Knollman, \pr {\bf 129}, 959 (1963).}
%
\bibitem{Jaksch}{D. Jaksch, C. Bruder, J.I.Cirac, C.W.Gardiner, and P. Zoller, \prl {\bf 81}, 3108 (1998).}
%
\bibitem{BHJoel}{G.J. Milburn, J.F. Corney, E.M. Wright and D.F. Walls, \pra {\bf 55}, 4318, (1997).}
%
\bibitem{Nemoto}{K. Nemoto, C.A. Holmes, G.J. Milburn, and W.J. Munro, \pra {\bf 63}, 013604 (2000).}
%
\bibitem{Penna}{V. Penna, \pre {\bf 87}, 052909 (2013).}
%
\bibitem{NDCJoel}{M.K. Olsen and J.F. Corney, \pra {\bf 94}, 033605 (2016).}
%
\bibitem{BECsplit}{C.V. Chianca and M.K. Olsen, \pra {\bf 92}, 043626 (2015).}
%
\bibitem{splitOC}{M.K. Olsen, \oc {\bf 371}, 1 (2016).}
%
\bibitem{ZhangECS}{Z. Jiang and C. M. Caves, \pra {\bf 93}, 033623 (2016).}
%
\bibitem{states}{M.K. Olsen and A.S. Bradley, \oc {\bf 282}, 3924 (2009).}
%
\bibitem{Pplus}{P.D. Drummond and C.W. Gardiner, \jpa {\bf 13}, 2353 (1980).}
%
\bibitem{Graham}{R. Graham, Springer Tracts Mod. Phys. {\bf }66, 1 (1973).}
%
\bibitem{Steel}{M.J. Steel, M.K. Olsen, L.I. Plimak, P.D. Drummond, S.M. Tan, M.J. Collett, D.F. Walls, and R. Graham, \pra {\bf 58}, 4824 (1998).}
%
\bibitem{BobWig}{R.J. Lewis-Swan, M.K. Olsen, and K.V. Kheruntsyan, \pra {\bf 94}, 033814 (2016).}
%
\bibitem{QNoise}{C.W. Gardiner and P. Zoller, {\em Quantum Noise}, (Springer-Verlag, Heidelberg, 2000).}
%
\bibitem{DFW}{D.F. Walls and G.J. Milburn, {\em Quantum Optics} (Springer-Verlag, Berlin, 1995).}
%
\bibitem{WigEPL}{L.I. Plimak, M.K. Olsen, M. Fleischhauer, and M.J. Collett, \epl {\bf 56}, 372 (2001).}
%
\bibitem{SMCrispin}{C.W. Gardiner, {\em Stochastic Methods: A Handbook for the Natural and Social Sciences}, (Springer-Verlag,
Berlin, 2002).}
\bibitem{Feshbach}{C. Chin, R. Grimm, P. Julienne, and E. Tiesinga, Rev. Mod. Phys. {\bf 82}, 1225 (2010).}
%
\bibitem{Albiez}{M. Albiez, R. Gati, J. F\"olling, S. Hunsmann, M. Cristiani, and M.K. Oberthaler, \prl {\bf 95}, 010402 (2005).}
%
\bibitem{phasegate}{M.K. Olsen and A.S. Bradley, \pra {\bf 91}, 043635 (2015).}
%
\bibitem{Bhattacharyya}{A. Bhattacharyya, Bull. Calcutta Math. Soc. {\bf 35}, 1 (1943).}
%
\bibitem{Greiner}{M. Greiner, O. Mandel, T. Esslinger, T. W. H\"ansch and I. Bloch, \nature {\bf 415}, 39-44 (2002).}

\end{thebibliography}
\end{document}